\documentclass[a4paper,superscriptaddress,twocolumn,prb,leqno]{revtex4-1}

\usepackage{float}
\usepackage{amsmath}
\usepackage{graphicx}
\usepackage{color}
\usepackage{epstopdf}

\usepackage{ulem}

\begin{document}
\title{THz conductivity of Sr$_{1-x}$Ca$_x$RuO$_3$}

\author{Diana Geiger}
\affiliation{1.\ Physikalisches Institut, Universit\"at Stuttgart, D-70569 Stuttgart, Germany}
\affiliation{Institut f\"ur Festk\"orperphysik, Technische Universit\"at Wien, A-1040 Vienna, Austria}
\author{Uwe S.\ Pracht}
\author{Martin Dressel}
\affiliation{1.\ Physikalisches Institut, Universit\"at Stuttgart, D-70569 Stuttgart, Germany}
\author{Jernej Mravlje}
\affiliation{Jozef Stefan Institute, SI-1000, Ljubljana, Slovenia}
\author{Melanie Schneider}
\author{Philipp Gegenwart}\thanks{present address: Experimentalphysik VI, Center for Electronic Correlations and Magnetism, Institute for Physics, Augsburg University, D-86135 Augsburg, Germany}
\affiliation{I.\ Physikalisches Institut, Georg-August-Universit\"at, D-37073 G\"ottingen, Germany}
\author{Marc Scheffler}
\email{marc.scheffler@pi1.physik.uni-stuttgart.de}
\affiliation{1.\ Physikalisches Institut, Universit\"at Stuttgart, D-70569 Stuttgart, Germany}
\email{scheffl@pi1.physik.uni-stuttgart.de}

\date{\today}

\begin{abstract}
We investigate the optical conductivity of Sr$_{1-x}$Ca$_x$RuO$_3$
across the ferromagnetic to paramagnetic transition that occurs at
$x=0.8$. The thin films were grown by metalorganic aerosol deposition 
with $0 \leq x \leq 1$ onto NdGaO$_3$ substrates. We performed THz 
frequency domain spectroscopy in a frequency range from 
3~cm$^{-1}$ to 40~cm$^{-1}$ (100~GHz to 1.4~THz) and at temperatures 
ranging from 5~K to 300~K, measuring transmittivity and phase shift 
through the films. From this we obtained real and imaginary parts of 
the optical conductivity.
The end-members, ferromagnetic SrRuO$_3$ and paramagnetic 
CaRuO$_3$, show a strongly frequency-dependent metallic response 
at temperatures below 20~K. Due to the high quality of these samples 
we can access pronounced intrinsic electronic contributions to the 
optical scattering rate, which at 1.4~THz exceeds the residual 
scattering rate by more than a factor of three. Deviations from a 
Drude response start at about 0.7~THz for both end-members in a 
remarkably similar way. 
For the intermediate members a higher residual scattering originating 
in the compositional disorder leads to a featureless optical response, 
instead. The relevance of low-lying 
interband transitions is addressed by a calculation of 
the optical conductivity within density functional theory in the local 
density approximation (LDA).
\end{abstract}


\maketitle
\section{Introduction}
  The optical response of a material provides access to its electronic
  behavior in a broad window of energy scales\cite{01basov11}. High-frequency ultraviolet and visible light spectroscopies probe interband transitions, 
  which give rise to an onset in the frequency-dependent optical absorption, 
  whereas infrared and THz spectroscopies at lower energies provide access to more subtle aspects
  of the collective low-frequency response of an electron gas. One of
  the most interesting aspects of the collective behavior are
  quantum phase transitions (QPTs)\cite{02VojtaQPT,03Loehn}. 
  Due to the absence of a characteristic scale near the
  QPT, one expects the optical response of such a material to be
  characterized by power laws. Power-law conductivity was found in several quantum-critical
  materials, such as MnSi\cite{03aMnSi,01basov11} and in cuprates close to optimal
  doping\cite{01basov11}. It was also observed in perovskite
  ruthenates\cite{04Kostic98,05Dodge00,06Lee02,07Kamal2006,01basov11}, the
  subject of the present paper.

The perovskite-structured ruthenate system Sr$_{1-x}$Ca$_x$RuO$_3$ is a
candidate material for a ferromagnetic QPT\cite{Cao1997, Khalifah2004, Fuchs2015}.  Its undoped parent
compound SrRuO$_3$ is an itinerant ferromagnet with ordering
temperature $T_{\textrm{C}}=160$~K and is of great interest for both
fundamental physics as well as applications\cite{08Koster2012}, whereas
the other parent compound, CaRuO$_3$, is a paramagnetic
metal. CaRuO$_3$ has been argued to be close to the magnetic critical
point as revealed by $T^{3/2}$ temperature dependence of dc resistivity\cite{09Capogna} below
30 K and logarithmic term in the temperature-dependent specific
heat\cite{10Cao08}. Very recently, the observation of Shubnikov-de-Haas
oscillations and $T^2$ resistivity below 1.5 K revealed that a fragile Fermi
liquid is recovered at low temperatures. In SrRuO$_3$ the Fermi liquid
is more robust with $T^2$ resistivity up to about
10 K\cite{09Capogna,11MackenzieSRO}.

THz and infrared optical studies on both compounds\cite{04Kostic98,05Dodge00,06Lee02,07Kamal2006} 
revealed an unusual optical response with the optical conductivity 
$\hat{\sigma}(\omega)=\sigma_1(\omega)+\mathrm{\textit{i}} \sigma_2(\omega)$ 
distinct from that of the Drude behavior that is given by
\begin{eqnarray}
\sigma_1(\omega)&=&\sigma_{\mathrm{dc}}\frac{1}{1+\omega^2\tau_D^2} \label{Drude} \\
\sigma_2(\omega)&=&\sigma_{\mathrm{dc}}\frac{\omega\tau_D}{1+\omega^2\tau_D^2}
\end{eqnarray}
with $\tau_D=1/\Gamma_D$ the Drude scattering time and
$\sigma_{\mathrm{dc}} = 1/\rho_{\mathrm{dc}}$ zero-frequency
conductivity\cite{12Dressel2002}. A recent optical
study on clean CaRuO$_3$ samples revealed Drude
behavior\cite{13GoettPRL2014} consistent with Fermi liquid concepts, but
only up to a frequency of 0.6 THz. Starting from 0.6 THz 
an abrupt increase of the optical scattering rate was found. A possible
origin of this behavior is the coupling to a critical low-energy 
fluctuation.

 A different point of view was offered in a very recent
 dynamical-mean-field theory study of CaRuO$_3$\cite{00Dang15},
 which pointed out the importance of low-lying interband transitions.
 Namely, the orthorhombic distortion of the cubic perovskite lattice
 leads to a series of mini-gaps (with the gap size of the order
 of THz), which significantly affect the optical response. 
 A related tetragonal ruthenate Sr$_2$RuO$_4$
 that occurs in a non-distorted structure does not show
 low-frequency deviations and exhibits a standard Fermi liquid optical
 response\cite{15Stricker14}, instead. One needs to add however, that
 Sr$_2$RuO$_4$ was also argued to be further away from the quantum
 critical point\cite{10Cao08}.

\begin{figure}[tbp]
  % Requires \usepackage{graphicx}
  \centering
  \includegraphics[width=0.5\textwidth]{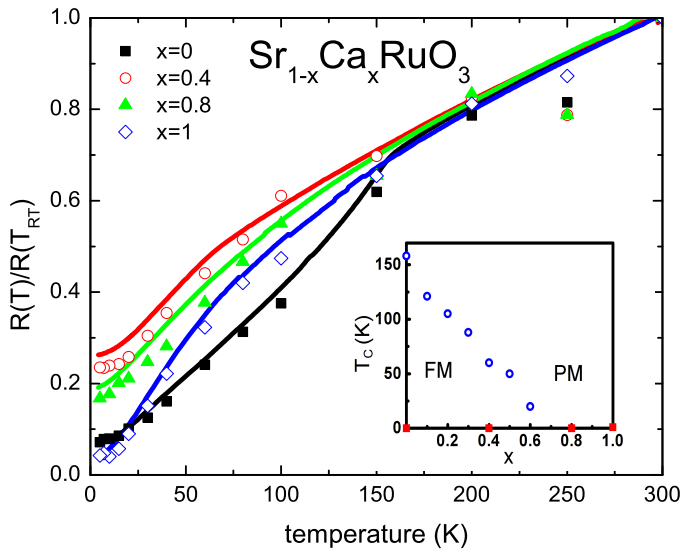}
  \caption{\small{(color online) Resistivity values, obtained from low-frequency 
  optical data (symbols) and four-point dc measurements 
(lines). Optical relative resistivity data were calculated from low-frequency 
(between 4~cm$^{-1}$ and 5~cm$^{-1}$) 
THz data by normalizing to the 300~K value, dc data are normalized to the 280~K value. 
Inset: phase diagram of Sr$_{1-x}$Ca$_x$RuO$_3$}. Circles represent Curie temperatures 
of differently doped thin-film samples\cite{16QPTSchneider}, 
squares indicate the doping levels that were investigated in this optical study.}
  \label{RRR}
\end{figure}

In the present work we address the evolution of the optical response of 
Sr$_{1-x}$Ca$_x$RuO$_3$ as a function of composition, across the magnetic 
phase transition. We have grown thin films, and for the compositions 
$x=0,x=0.4,x=0.8$ and $x=1$ (denoted by solid symbols on the phase diagram 
in the inset of Fig.~\ref{RRR}) the THz response was measured. 
The end-members of the series are sufficiently clean to exhibit optical 
conductivity that corresponds to a strongly frequency-dependent scattering 
rate. We investigate the optical response also theoretically within 
density functional theory in the local-density approximation. The calculated 
optical scattering rate for both CaRuO$_3$ and SrRuO$_3$ shows strong frequency 
dependence even though frequency-independent scattering is put into the 
calculation; this suggests the relevance of low-lying interband transitions.

The paper is structured as follows: in Sec.~\ref{sec:exp} we present
the details of the sample growth and the experimental setup. In
Sec.~\ref{sec:res} we show the measured optical conductivities and the
associated optical scattering rate. In Sec.~\ref{sec:theory} we present the optical
conductivity as calculated within the LDA approximation and discuss
the relevance of the interband transitions with respect to the experimental
results. In Sec.~\ref{sec:summary} we summarize and conclude, and in the Appendix
%Sec.~\ref{sec:scale} 
we analyze the scaling properties of the measured conductivities.  

\section{Samples and Experimental Setup} \label{sec:exp}

\begin{figure}[tbp]
  % Requires \usepackage{graphicx}
  \centering
  \includegraphics[width=0.5\textwidth]{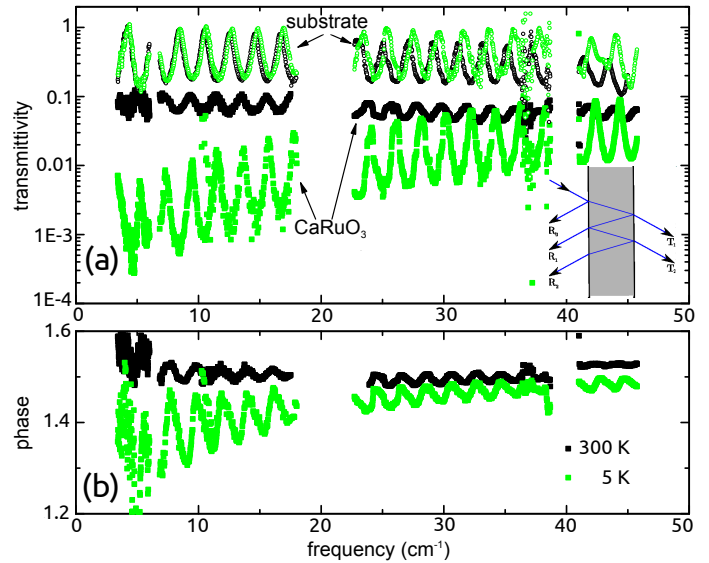}
  \caption{\small{(color online) Transmittivity (a) and phase (b) data of CaRuO$_3$ on 
  NdGaO$_3$ and bare NdGaO$_3$ substrate at 300~K and 5~K. Open
  circles correspond to substrate data, 
  full squares to thin-film samples. Inset of (a): multiple reflections in a dielectric 
  substrate lead to Fabry-P\'{e}rot oscillations in the raw data.}}
  \label{SigmaExamples}
\end{figure}

The epitaxial Sr$_{1-x}$Ca$_x$RuO$_3$ thin films were grown in G\"ottingen by 
metalorganic aerosol deposition (MAD) \cite{17Moshny,18AerosolMelanie}. We investigated 
the four different compositions SrRuO$_3$, Sr$_{0.6}$Ca$_{0.4}$RuO$_3$, 
Sr$_{0.2}$Ca$_{0.8}$RuO$_3$, and CaRuO$_3$, all deposited on NdGaO$_3$(110) substrates. 
The thin films grow epitaxially on the substrate with the same [110] orientation. 
NdGaO$_3$ is the substrate of choice:  its lattice constant matches the thin films well, 
and its dielectric properties are convenient for transmittivity
measurements. More precisely, for SrRuO$_3$ the substrate leads to 1.7\% of compressive 
strain\cite{18aSROstrain,18bSROstrain,LuSciRep15} and for CaRuO$_3$ to about -0.4\% 
of tensile strain \cite{08Koster2012,18cCROstrain,13GoettPRL2014}. Our measurements of the 
lattice constant perpendicular to the substrate are consistent with the results 
referenced above (not shown).
Another popular substrate, SrTiO$_3$ has a very large frequency- and temperature-dependent 
dielectric function, which makes optical measurements difficult\cite{19SCESProceeding,20Felger2013}.
\begin{table}
\begin{tabular}{|c|c|c|c|c|c|}
\hline 
$x$&thickness(nm)&$T_\mathrm{C}$(K)&RRR&$\rho_\mathrm{dc}$($\mu\Omega$cm)&$\rho_\mathrm{THz}$($\mu\Omega$cm)\\
\hline
0 & 70 & 150 & 15 & 194 & 185 \\
0.4 & 80 & 60 & 3.8 & 259 & 290 \\
0.8 & 77 & - & 5.4 & 215 & 230 \\
1 & 40 & - & 31 & 230 & 237 \\
\hline 
\end{tabular}
\caption{Properties of the studied thin-film samples. Depending on calcium content $x$, the table shows: 
the film thickness, the Curie temperature $T_\mathrm{C}$ determined from dc resistivity measurements (derivative), 
the residual resistivity ratio RRR from $R$(300\,K)$/R$(0\,K) with a low-temperature extrapolation for the 0\,K value, 
the dc resistivity value at 280\,K and the low-frequency THz resistivity value at 300\,K (both used in the 
normalized curves of Fig.\ \ref{SigmaExamples}).}
\label{TableSamples}
\end{table}
Some relevant sample properties are listed in table \ref{TableSamples}. All four samples are of high 
quality and clearly metallic, as evident from the resistivity curves and residual resistivity ratios 
(RRR) between 300 K and 0 K (extrapolated value) in Fig.\ \ref{RRR} and table \ref{TableSamples}. 
While the SrRuO$_3$ and CaRuO$_3$ have high RRRs of 15 and 31, respectively, the higher disorder in 
the $x=0.4$ and $x=0.8$ samples enhances the scattering and reduces the RRR to 4 and 5.  

The transmittivity and phase response of all samples was investigated in Stuttgart in a THz frequency domain 
spectrometer with a Mach-Zehnder interferometer\cite{21THz,22PrachtIEEE}. 
Fig.\ \ref{SigmaExamples} shows raw data of CaRuO$_3$ at 300~K and 5~K; for comparison, the transmittivity
raw data of an empty NdGaO$_3$ substrate are also plotted. A frequency range from around 
3~cm$^{-1}$ to 40~cm$^{-1}$ was covered by measurements with five different backward wave oscillator sources. 
The prominent Fabry-P\'{e}rot type oscillations in the transmitted signal are typical to the raw data of 
metallic thin films on dielectric substrates and caused by the substrate, as indicated 
schematically in the inset of Fig.\ \ref{SigmaExamples}. The contribution of the film is a modulation of 
these substrate resonances: 
the difference between the data of bare NdGaO$_3$ and CaRuO$_3$ on NdGaO$_3$ in Fig.\ \ref{SigmaExamples}(a)
is caused by the film. While the pure NdGaO$_3$ spectra are almost temperature 
independent, there is a strong temperature dependence in the thin-film samples. 
Already in the raw data we see a considerably lower transmittivity at 5~K than at 300~K due to an increased 
conductivity in the metallic film. 
The optical conductivity is evaluated directly from the raw data: since both transmittivity and phase are measured, 
no Kramers-Kronig analysis is required. Each Fabry-P\'{e}rot maximum 
is analyzed to obtain the complex conductivity $\sigma_1$ and $\sigma_2$ at its peak frequency.\cite{12Dressel2002,22PrachtIEEE} 
From these values, a frequency-dependent 
scattering rate $\Gamma(\omega)$ can be calculated using the extended Drude formalism:
\begin{equation}\label{EqExtendedDrude}  
\frac{\Gamma(\omega)}{\omega_{\mathrm{P}}^2} = \frac{\rho_1(\omega)}{4\pi} = \frac{1}{4\pi} \frac{\sigma_1(\omega)}{{\sigma_1(\omega)}^2+{\sigma_2(\omega)}^2}
\end{equation}
with ${\omega_{\mathrm{P}}}^2=4\pi n e^2/m$ being the plasma frequency, $e$ and $m$ the (free-)electron charge 
and mass and $n$ the charge carrier density.\cite{12Dressel2002,22a-ex37Scheffler2013}

\section{Results}  \label{sec:res}
\begin{figure*}[tbp]
  % Requires \usepackage{graphicx}
  \centering
  \includegraphics[width=\textwidth]{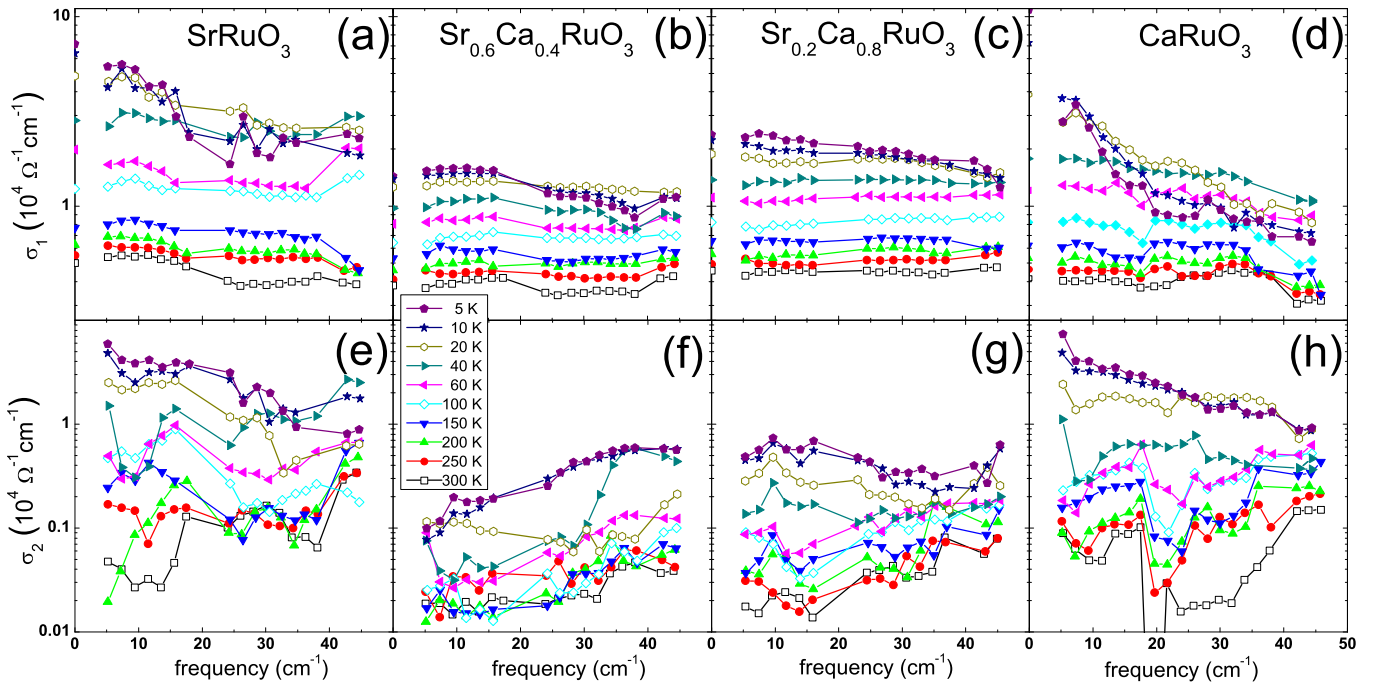}
  \caption{\small{(Color online) Real and imaginary parts of the optical conductivity of the four studied Sr$_{1-x}$Ca$_x$RuO$_3$ compositions at various 
  temperatures between 300~K and 5~K.\cite{Footnote}}}
  \label{Sigma1}
\end{figure*} 
Fig.\ \ref{Sigma1} shows the frequency-dependent real and imaginary
parts of the conductivity of all samples. The absolute values of
$\sigma_1$ are characteristic of metals and of the order of 
10$^4$~$\Omega^{-1}$~cm$^{-1}$. At temperatures above 100~K, all samples
show nearly frequency-independent real and low imaginary parts: the
scattering rate exceeds our measured frequency range.
A maximum in $\sigma_1(\omega)$ has been observed in the mid-infrared at these
temperatures. \cite{01basov11,06Lee02} In our data there is no visible 
onset of this maximum, $\sigma_1(\omega)$ drops monotonically.\cite{Footnote}
\begin{figure}[tbp]
  % Requires \usepackage{graphicx}
  \centering
  \includegraphics[width=\columnwidth]{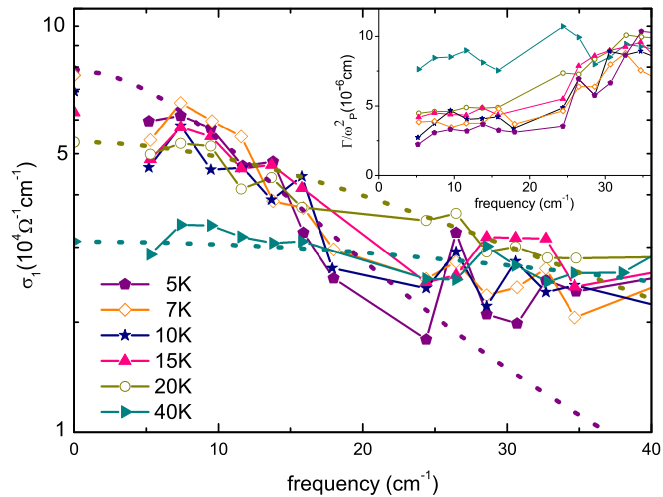}
  \caption{\small{(Color online) $\sigma_1(\omega)$ of SrRuO$_3$ for low temperatures. The dotted lines at 5 K, 20 K, 
  and 40 K are Drude fits following eq.\ \ref{Drude}} with $\sigma_{\mathrm{dc}}$ values fixed from dc measurements. 
  All points were taken into account for the 20 K and 40 K fits, the 5 K fit includes only frequencies below 30 cm$^{-1}$ 
  due to the non-Drude-like plateau at higher frequencies. Inset: Scattering rates for corresponding temperatures.}
  \label{DrudeFitsSRO}
\end{figure}
In Fig.\ \ref{RRR} we compare the THz response at our low-frequency limit (in particular the real part $\rho_1$ of the frequency-dependent resistivity 
$\hat{\rho}(\omega) = \rho_1(\omega) + i \rho_2(\omega) = 1/\hat{\sigma}(\omega)$) with results from dc measurements. There is very good agreement between 
these two data sets. This suggests that at least in this temperature range there are no additional features in the optical conductivity at lower frequencies.
Considering that dc resistivity current paths and THz beam spots probe do not necessarily probe the same sample areas, Fig.\ \ref{RRR} also documents the homogeneity of the films. 
\subsection{SrRuO$_3$}
The $\sigma_1$ and $\sigma_2$ spectra of SrRuO$_3$ for exemplary temperatures are shown in Fig.\ \ref{Sigma1} (a) and (e), respectively. 
The $\sigma_1$ spectrum, rather flat at high temperatures, changes upon cooling: below 40~K, 
$\sigma_1(\omega)$ develops a drop towards higher frequencies that moves to lower frequencies, down to around 12~cm$^{-1}$ with decreasing temperature.
In terms of the Drude model the drop indicates the scattering rate $\Gamma/2\pi$. 
To be consistent with a Drude interpretation, $\sigma_2(\omega)$ should show a maximum at the corresponding frequency, passing through our spectral range as a function 
of temperature\cite{Dressel2006} and leaving monotonous frequency-dependent behavior in the limiting cases, namely increasing for high temperatures and decreasing for low temperatures, as found in our data. To address the low-temperature properties of SrRuO$_3$ in more detail, Fig.\ \ref{DrudeFitsSRO} shows $\sigma_1(\omega)$ for all measured temperatures below 40~K. In addition, we show Drude fits following eq.\ \ref{Drude} for a few temperatures.
\begin{figure}[tbp]
  % Requires \usepackage{graphicx}
  \centering
  \includegraphics[width=\columnwidth]{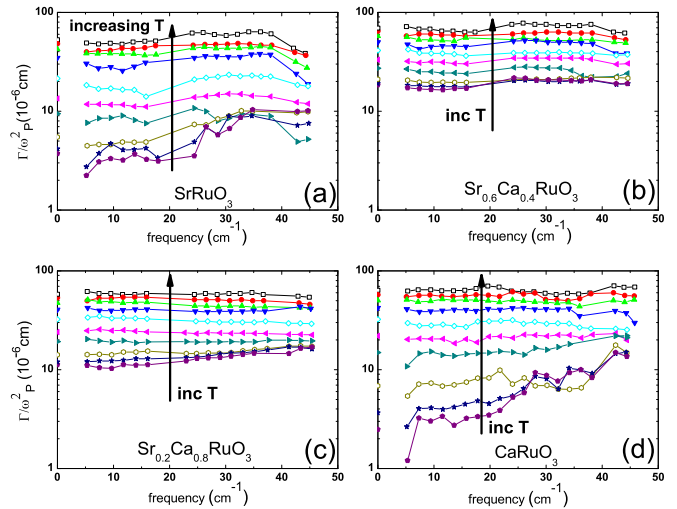}
  \caption{\small{(Color online) Frequency-dependent scattering rates at 300~K, 250~K, 200~K, 150~K, 100~K, 60~K, 40~K, 20~K, 10~K, and 5~K. Colors and 
  symbols identical to Fig. \ref{Sigma1}.\cite{Footnote2}}}
  \label{Scattering}
\end{figure}
\begin{figure}[tbp]
  % Requires \usepackage{graphicx}
  \centering
  \includegraphics[width=\columnwidth]{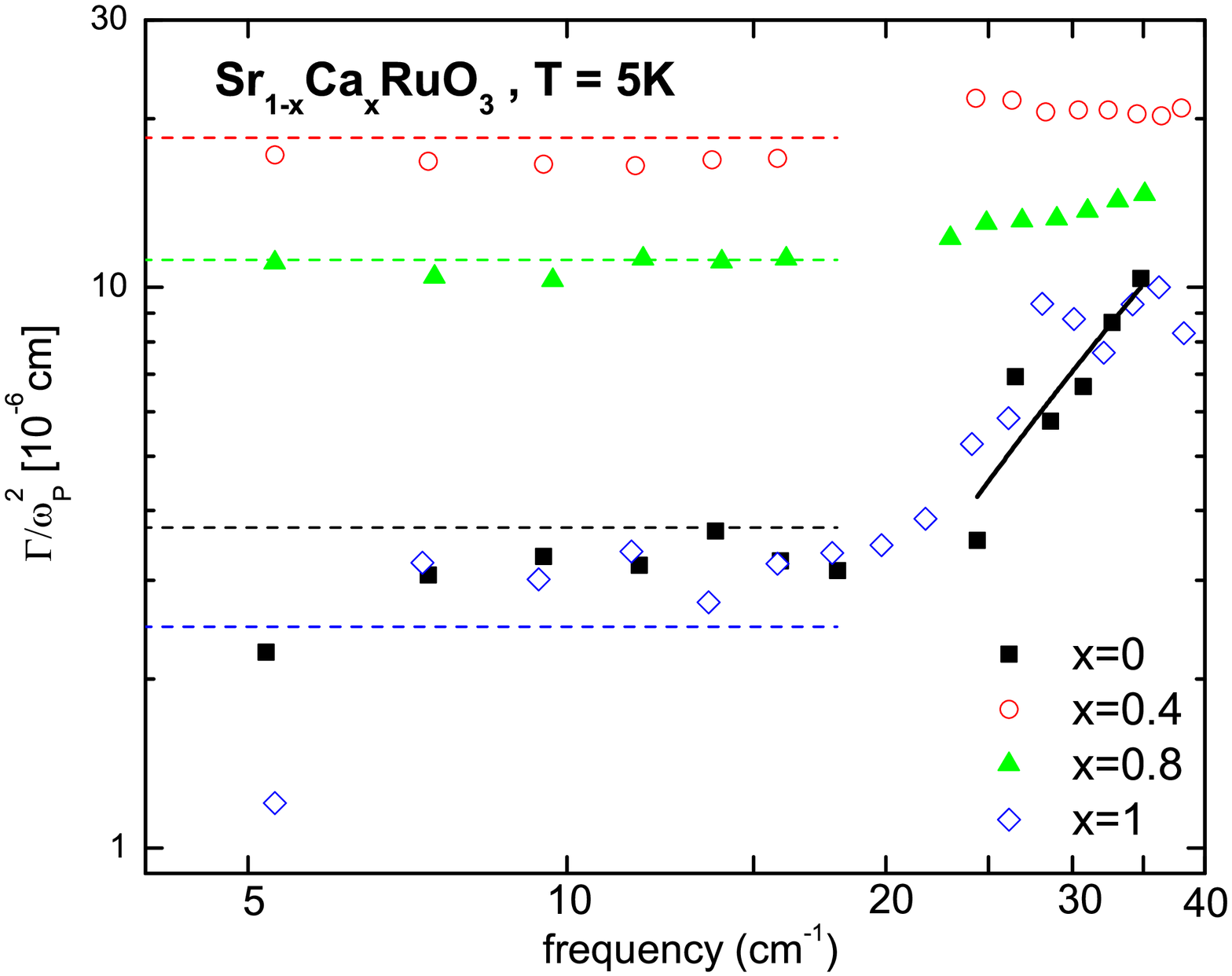}
  \caption{\small{(Color online) Scattering rates of Sr$_{1-x}$Ca$_x$RuO$_3$ at temperature 5~K. The dashed guides indicate the value 
  of the dc scattering rate, which is proportional to 1/$\sigma_{\mathrm{dc}}$. 
  The black straight line is the best FL fit, $\Gamma = \Gamma_0 + a\, \omega^2$ to the high-frequency range of the SrRuO$_3$ above 20 cm$^{-1}$.}}
  \label{ScatteringLT}
\end{figure}
Within the measurement accuracy, the Drude fits describe the conductivity data well above 20~K. However, at lower temperatures, the conductivity of SrRuO$_3$ deviates from the Drude fit because at high frequencies $\sigma_1(\omega)$ is not reduced further but remains constant.

This plateau for frequencies above 20~cm$^{-1}$ is present in all spectra of Fig.\ \ref{DrudeFitsSRO} and becomes more evident with decreasing temperature. Since the plateau cannot 
be described as part of a single Drude term, the fit at 5 K only includes the points below 30 cm$^{-1}$. 
For further analysis we use the extended Drude formalism, eq.\ \ref{EqExtendedDrude}. 
The obtained scattering rate is shown in Fig.\ \ref{Scattering}(a) for the same temperatures as in Fig.\ \ref{Sigma1}. 
Above 40~K the scattering rate is basically constant as a function of frequency, corresponding to a simple Drude behavior.
Below 20~K, Fig.\ \ref{Scattering}(a) indicates a pronounced frequency dependence of the scattering rate, an increase towards higher frequencies.

Deviations from Drude behavior were noticed earlier by Dodge \textit{et al} \cite{05Dodge00}.
Our $\sigma_1(\omega)$ of SrRuO$_3$ qualitatively follows this: flat for high temperatures with a broad low-frequency peak emerging at lowest temperatures.
The high quality of our samples, achieved with MAD \cite{18AerosolMelanie}, leads to 2-3 times higher absolute values of $\sigma_1$ at low temperatures compared to earlier experiments 
\cite{05Dodge00}, which enables us to charactrerize those deviations in better detail. 

For this purpose we turn to Fig.\ \ref{ScatteringLT}, where the scattering rate at 5~K is plotted. It is mostly frequency independent below 20~cm$^{-1}$, in the range below the plateau in $\sigma_1(\omega)$. Above 20~cm$^{-1}$ it increases strongly with frequency, approximately by a factor of three between 20~cm$^{-1}$ and 35~cm$^{-1}$. 
This increase roughly corresponds to a quadratic FL-like frequency dependence, as indicated by the straight line in Fig.\ \ref{ScatteringLT}. However, the transition from a flat to 
a steep slope in this double-logarithmic plot is much too abrupt to fit the FL prediction $\Gamma(\omega) = \Gamma(\omega=0) + b (\hbar \omega)^2$ in the entire frequency range.
Clearly, simple Fermi-liquid considerations are not sufficient to explain the observed low-temperature frequency dependence of the scattering rate. The low-lying interband transitions\cite{00Dang15} that can reproduce such a steep onset of optical scattering rate are  discussed in Sec.~\ref{sec:theory}.

\subsection{Sr$_{0.6}$Ca$_{0.4}$RuO$_3$ and Sr$_{0.2}$Ca$_{0.8}$RuO$_3$}
The conductivity spectra of the doped samples are shown in Fig.\ \ref{Sigma1} (b), (c), (f), and (g). For both compositions, $\sigma_1(\omega)$ is almost constant, and $\sigma_2$ is much smaller than $\sigma_1$; this suggests a conventional metallic conductivity where the scattering rate is much higher than the studied spectral range. A weak frequency dependence only emerges at the lowest temperatures. Upon cooling, both $\sigma_1(\omega)$ and $\sigma_2(\omega)$ increase, which is consistent with a decrease in scattering rate. From the comparison of all samples 
in Fig.\ \ref{Sigma1} we find that the conductivity of the doped samples is substantially lower than of the undoped ones. 
This we attribute to compositional disorder, which significantly increases the residual scattering. As expected, this effect is more pronounced for Sr$_{0.6}$Ca$_{0.4}$RuO$_3$, where 
the disorder due to the composition will be particularly strong.
This scenario is consistent with the comparatively small increase of the conductivity upon cooling, and correspondingly with the small residual resistivity ratio. 
Unlike in the undoped samples there is no strong roll-off in $\sigma_1(\omega)$ at low frequencies. The extended Drude analysis yields a mostly frequency-independent scattering rate 
(Fig.\ \ref{Scattering}), substantially higher at low temperatures than in the pure compounds (Fig.\ \ref{ScatteringLT}).
\begin{figure}[tbp]
  % Requires \usepackage{graphicx}
  \centering
  \includegraphics[width=\columnwidth]{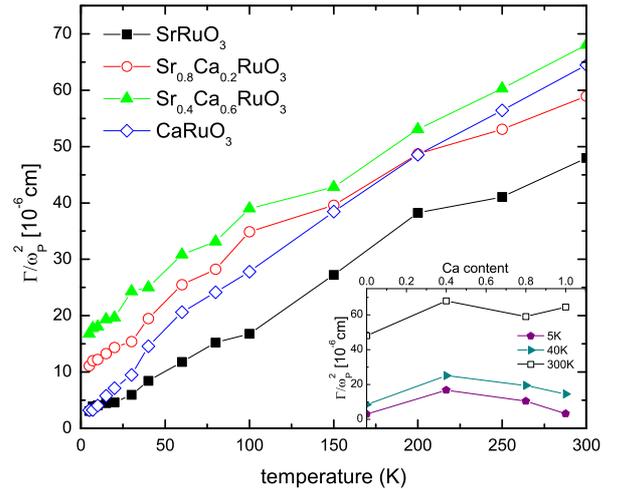}
  \caption{\small{(Color online) Scattering rate $\Gamma$  from extended Drude analysis at frequency 7~cm$^{-1}$ for the four different compositions Sr$_{1-x}$Ca$_x$RuO$_3$ as a 
   function of temperature (main plot) and as a function of Ca content $x$ (inset).}}
  \label{ScatteringRateVsTemperature}
\end{figure}
Therefore the strong disorder scattering in the doped samples dominates their transport properties at low temperatures, which makes it impossible to deduce signatures of electronic 
correlations from the THz data at this stage. This is unfortunate, as Sr$_{0.2}$Ca$_{0.8}$RuO$_3$ is located close to the QPT from ferromagnetism to paramagnetism, and here the influence 
of quantum-critical fluctuations to the electronic scattering would be of particular interest.

That the high scattering rate at low temperatures in these two samples indeed stems from the disorder is illustrated by a comparison with the pure compounds in 
Fig.\ \ref{ScatteringRateVsTemperature}. The main plot shows the scattering rate as a function of temperature, determined from the extended Drude analysis at 7~cm$^{-1}$, for all
samples of this study. The scattering rates of Sr$_{0.6}$Ca$_{0.4}$RuO$_3$ and Sr$_{0.2}$Ca$_{0.8}$RuO$_3$ basically are shifted to higher values when compared to SrRuO$_3$, and this 
temperature-independent offset is stronger for Sr$_{0.6}$Ca$_{0.4}$RuO$_3$. The inset of Fig.\ \ref{ScatteringRateVsTemperature} shows the scattering rate as a function of Ca content 
$x$ for three exemplary temperatures. At the lower temperatures, the dome-shape dependence with minimal scattering for the pure compounds at the two outer ends and a maximum around the 
center is evident.

\subsection{CaRuO$_3$} 
\begin{figure}[tbp]
  % Requires \usepackage{graphicx}
  \centering
  \includegraphics[width=\columnwidth]{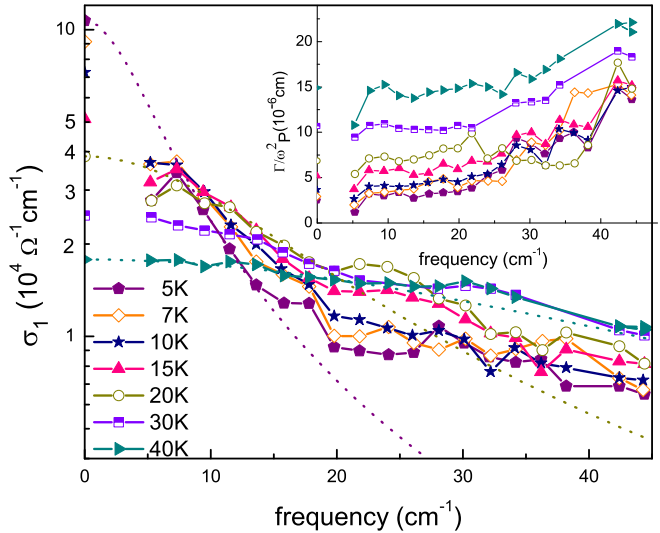}
  \caption{(Color online) $\sigma_1$ spectra of CaRuO$_3$ for low temperatures with tentative Drude fits. Due to the high-frequency plateau, the fits do not work at lowest temperatures. 
  Inset: Scattering rates for corresponding temperatures.   \label{CROLowT}}

\end{figure}

CaRuO$_3$ exhibits a similar behavior as SrRuO$_3$: at high temperatures, $\sigma_1(\omega)$ is almost constant throughout our spectral range and $\sigma_2(\omega)$ is very small 
(see Fig.\ \ref{Sigma1}(d) and (h)). 
This is consistent with a frequency-independent scattering rate as shown in Fig.\ \ref{Scattering}(d). Upon cooling, $\sigma_1$ and $\sigma_2$ continuously increase; down to approximately 
40~K the frequency dependence of $\sigma_1(\omega)$ as well as $\Gamma(\omega)$ is weak. This changes at lower temperatures (see Fig.\ \ref{CROLowT}): at frequencies below 
10~cm$^{-1}$, $\sigma_1$ strongly increases while at higher frequencies it is rather constant or even slightly decreasing.
At the lowest temperatures, e.g. 5~K, this leads to a pronounced roll-off in $\sigma_1(\omega)$ between 5~cm$^{-1}$ and 10~cm$^{-1}$, followed by a non-Drude-like plateau. 
The frequency-dependent scattering rate, extracted from the extended Drude analysis, is shown in in Fig.\ \ref{Scattering}(d), and the 5~K one is compared to the other compositions in 
Fig.\ \ref{ScatteringLT}.

The optical data on CaRuO$_3$ was discussed earlier by Schneider \textit{et al.}, \cite{13GoettPRL2014} and it was found to be inconsistent with Drude or FL behavior \cite{25Berthod2013} 
above 15~cm$^{-1}$ or so. The deviations from Drude behavior are displayed also on Fig.~\ref{CROLowT}. The high quality of the CaRuO$_3$ sample is evident from the low scattering rate at 
lowest temperatures, as clearly discernible from the comparison with the other samples in Fig.\ \ref{ScatteringRateVsTemperature}. With increasing temperature, this low-frequency scattering 
rate increases more strongly than for SrRuO$_3$ (and the intermediate samples) due to stronger electronic correlations in the absence of ferromagnetism\cite{Dang_PRB2015}.

% Previous studies \cite{16QPTSchneider,26QPTYoshimura} have already shown that there is NFL behavior in the dc resistivity of CaRuO$_3$ between 2~K and 10~K, where 
% $(\rho_{\mathrm{dc}}(T) - \rho_0) \propto T^\alpha$ with $\alpha \approx 1.5 < 2$ was found\cite{16QPTSchneider}.
In previous optical studies on CaRuO$_3$, Lee
\textit{et al.} \ mainly focused on data above 40~cm$^{-1}$ in a large
temperature range without measuring many low temperatures
\cite{06Lee02}. This study has substantial overlap in temperature and
frequency range with Kamal \textit{et al.} \cite{07Kamal2006}. Our
conductivity spectra match these two previous works well and are
consistent if quantitative shifts due to our enhanced film quality are taken into
account.  In our study, the most pronounced frequency and temperature
dependence of the conductivity is located at the lowest energies:
temperatures below 20~K and frequencies below 10~cm$^{-1}$, beyond the
limits of the previous works.  Scaling plots were proposed and were
controversially discussed in Lee \textit{et al.}\cite{06Lee02} and
Kamal \textit{et al.}\cite{07Kamal2006}. To compare with these earlier
works, we replotted our data in the suggested form, see Appendix. Our
data on SrRuO$_3$ does not scale well.  For CaRuO$_3$, the scaling
seems to be obeyed, but as was discussed in Ref.~\onlinecite{07Kamal2006},
the accessible frequency and temperature ranges are too narrow and the
noise level too high to draw any firm conclusions.

\section{Low frequency optical conductivity within LDA} \label{sec:theory}
Very recently it was argued that low-lying optical interband
transitions, which are activated by orthorhombic distortions, might
affect the optical response at unexpectedly low frequencies. In
particular, a dynamical mean-field theory (DMFT) study\cite{00Dang15} argued that the low
frequency deviation from Drude optics in CaRuO$_3$ occurs already at
the level of band theory. As a test for the effects of the band
structure in the simplest possible setting, we calculated the optical
conductivity of SrRuO$_3$ and CaRuO$_3$ (in their bulk structure)
using LDA as implemented in the Wien2k
package. \cite{29Wien2k,30Draxl06} In this approach, the current
matrix elements are evaluated from the band structure and the optical
conductivity is evaluated for a frequency-independent scattering rate
as described by a self energy $\Sigma=-i (1/\tau_D)/2$. Similar
approach was used in iron-based superconductors\cite{30_boeri,30_calderon}
where also interband contributions were found to be important.

\begin{figure}[tbp]
  % Requires \usepackage{graphicx}
  \centering
  \includegraphics[width=0.5\textwidth]{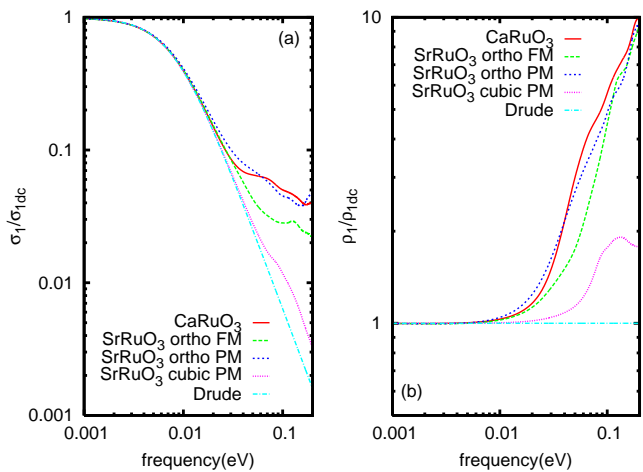}
  \caption{\small{(Color online)(left) Optical conductivity in the LDA approximation with static impurity scattering $1/\tau_D=0.008$~eV for CaRuO$_3$ and for SrRuO$_3$ compared to the Drude 
  optical conductivity. For SrRuO$_3$ besides ferromagnetic calculation in an orthorombic structure also the paramagnetic result in the orthorombic and cubic structure are shown. (right) The 
  corresponding optical scattering rates, normalized to the dc value.}}
  \label{plot_lda}
\end{figure}

The frequency-dependent real part of the optical conductivity,
normalized to the zero-frequency value, is displayed in Fig.\ref{plot_lda}.
One sees that the calculated optical conductivities for orthorhombic 
structures deviate from the Drude behavior
$\sigma_1(\omega)/\sigma_\mathrm{dc}=1/(1+\omega^2 \tau_D^2)$ at very low
frequencies of the order of 20~meV. These deviations originate in low-lying interband 
transitions across mini-gaps that are opened up by
orthorhombic distortions. Nonmagnetic SrRuO$_3$ shows very similar 
behavior to CaRuO$_3$ but with obvious deviations from Drude
behavior at even lower frequencies, which is due to the fact
that the orthorhombic distortion is smaller there. The shift of the bands
due to the exchange splitting in ferromagnetic SrRuO$_3$
(with moment 1.6~$\mu_B$) diminishes these deviations. The calculated
optical conductivity in the cubic structure deviates from the Drude
behavior starting at a higher frequency and in a much less pronounced way.

For easier comparison with the experimental data, we extracted also
$\rho_1(\omega)=\mathrm{Re} (1/\hat{\sigma}(\omega))$, which is proportional to the optical
scattering rate as defined by the extended Drude formalism. This presents a very clear influence of the
band-structure effect. Although the actual scattering rate put into the
calculation is frequency independent, the inferred optical scattering rate
(that attempts to describe the response of a multi-band system in terms
of the response of a single-band one) exhibits a strong frequency dependence.

In these calculations, SrRuO$_3$ and CaRuO$_3$ show a deviation
from the Drude dynamics at a similar frequency of about $20$~meV. To
compare this value to the frequency in the experiment, one needs to
additionally divide the value by the corresponding renormalizations.  
In CaRuO$_3$, the measured specific heat is about 7 times above the 
value found in the band theory.\cite{41CROCp} Therefore the expected 
frequency for the deviation from Drude dynamics is 20~meV/7~$\approx$~3 meV
($\approx$~0.7~THz~$\approx$~24~cm$^{-1}$) and consistent with the experiment.\\

 In SrRuO$_3$ the renormalization is smaller, about 4 judging from the
 specific heat.\cite{42SROCp} Furthermore, mostly the
 minority carriers (with the plasma frequency
 $\omega_{p\downarrow}\sim 2.6$~eV) and not the majority carriers (that
 correspond to small Fermi surfaces of the almost completely filled
 bands and the plasma frequency $\omega_{p\uparrow} \sim 1.3$~eV) contribute
 to the conductivity and within LDA+DMFT it was found\cite{00Dang15} 
 that the renormalization for these
 minority carriers is smaller, only about 2.

From these considerations, the deviation from Drude behavior through
band structure effects would only be expected at a frequency scale
above our experimental range. However, one needs to keep in mind that
the growth of SrRuO$_3$ on NdGaO$_3$ substrate modifies the structure 
substantially\cite{18aSROstrain,18bSROstrain},
which may decrease the corresponding scale. This should be tested in future
theoretical calculations or experimentally by growth on different substrates.

\section{Conclusions and outlook} \label{sec:summary}

The optical conductivity of the four samples SrRuO$_3$,
Sr$_{0.6}$Ca$_{0.4}$RuO$_3$, Sr$_{0.2}$Ca$_{0.8}$RuO$_3$, and
CaRuO$_3$ of the material system Sr$_{1-x}$Ca$_x$RuO$_3$ with a QPT at
$x\approx 0.8$ were investigated with THz spectroscopy.  The
transmittivity and phase data revealed metallic behavior for all
samples. The doped samples have a comparably high scattering rate due
to compositional disorder, and their THz properties can be described
within a simple Drude picture.  The SrRuO$_3$ and CaRuO$_3$ samples,
in contrast, at low temperatures have low scattering rates
consistent with the high RRRs that could be achieved by MAD growth. At
low temperatures, SrRuO$_3$ and CaRuO$_3$ can be described by a Drude
response with constant scattering rate only up to approximately
20~cm$^{-1}$. By calculating the optical response within the band
theory we have shown that the deviations from Drude behavior may be
caused by low-lying interband transitions. Especially for CaRuO$_3$
the calculated response is very similar to the measured one, whereas
for SrRuO$_3$ the calculated scattering rate deviates at a frequency
that is too high to account for our measurements. A possible origin of
this discrepancy is a deviation of the thin-film structure from the
bulk-one for which the calculations were made, an issue that needs to
be explored in future work.

On the experimental side, future studies should on the one hand
attempt to reach lower temperatures \cite{22PrachtIEEE,40Bachar2015,Basov3He}
and frequencies \cite{31Scheffler2005a,32Steinberg2012} in particular
to check if the crossover to a scattering rate quadratic in frequency
that would indicate the elusive FL optical response
\cite{33Dressel2011,34Nagel2011,35Chubukov2012,36Maslov2012,22a-ex37Scheffler2013}
can be found. Furthermore, optical measurements on our MAD-grown
samples at higher frequencies, in the infrared range, should be
performed to allow direct comparison with previous studies of samples
with somewhat lower RRR at infrared frequencies and to further
investigate the nature of the non-Drude plateau. At these larger
frequencies not only the band-effects but also the genuine
contribution from correlations~\cite{00Dang15} will contribute.

Further improvement in sample growth can be envisaged, too. While the
scattering rate for doped samples with random arrangement of Sr and
Ca atoms will remain too high to reveal interesting behavior in our
frequency range, further improvements of the residual scattering of
CaRuO$_3$ thin films have recently been achieved by growth on
NdGaO$_3$ substrates with vicinal cut \cite{13GoettPRL2014}.  THz
studies on such samples might be more difficult due to birefringence
in the substrate,\cite{38Scheffler2009,39Ostertag2011} but finally
such studies might reveal even more clearly the unconventional
scattering rate of CaRuO$_3$ at THz frequencies.

\section{Appendix - Scaling analysis} \label{sec:scale}
\begin{figure}[tbp]
  \centering
  \includegraphics[width=0.49\textwidth]{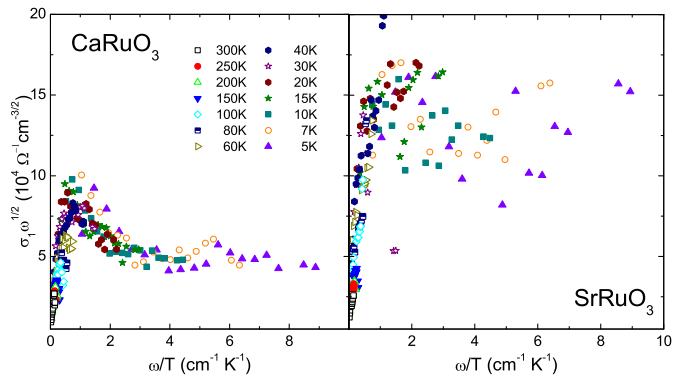}
  \caption{$\omega$/T scaling plots for SrRuO$_3$ and CaRuO$_3$ following Refs.\ \onlinecite{07Kamal2006} and \onlinecite{06Lee02}}.
  \label{FigScaling}
\end{figure}

Previous optical studies found that $\sigma_1(\omega,T)$ data for
SrRuO$_3$ and CaRuO$_3$ can be combined into a scaling plot following
eq.
\begin{equation}
\sigma_1(\omega, T) = \omega^{-1/2} Z(\omega/T)\label{scaling}
\end{equation}
with scaling function $Z(\omega/T)$.\cite{07Kamal2006,06Lee02} Such scaling of an experimentally accessible response can be a strong indication for quantum-critical 
behavior,\cite{27Schroeder2000,28Stockert2010} but the initial suggestion\cite{06Lee02} that the apparent scaling of $\sigma_1(\omega,T)$ in SrRuO$_3$ and CaRuO$_3$ should be interpreted as quantum 
critical was later rebutted.\cite{07Kamal2006} To present our data in the context of the scaling analysis of those previous works, we plot $\sigma_1(\omega,T) \omega^{1/2}$ vs.\ $\omega/T$ 
in Fig.\ \ref{FigScaling}, which should lead to data collapse if eq.\ \ref{scaling} were appropriate to describe the data. While our high-temperature spectra collapse into a small region of the 
scaling plot for both samples, below 40~K the samples differ strongly: in the case of CaRuO$_3$, within the experimental error, all temperatures collapse to one scaling curve, which slightly broadens 
for lower temperatures. For SrRuO$_3$ the spectra fail to collapse below 40~K. Although CaRuO$_3$ and SrRuO$_3$ show similar features in their optical conductivities and 
scattering rates, they are different with respect to the low-energy scaling behavior. Our low-frequency and low-temperature spectra of SrRuO$_3$ cannot be scaled by expression \ref{scaling}, while it 
works well for CaRuO$_3$.


\begin{thebibliography}{revtex4}
\bibitem{Footnote} Our measured data start with dc measurements ($f = \omega /2\pi = 0$), followed by frequencies from around 5 cm$^{-1}$ to 
40 cm$^{-1}$. We find that at the lowest frequencies it becomes increasingly difficult to obtain precise values for the concuctivity 
because of stray radiation: a fraction of the comparatively broad THz beam can reach the detector without interfering with the sample 
and therefore cause a higher experimentally determined transmittivity. This becomes most evident at high conductivity values because the 
absolute signal strength is already very low. In our measurements with highest $\sigma_1$ (found in the undoped samples at low temperatures) 
the lowest frequency point is lower than expected. We attribute this feature to stray radiation and do not interpret it in terms of a 
physical property of the sample.
\bibitem{Footnote2} At the highest frequencies, correspondingly at the shortest wavelength, the THz phase measurement is more difficult, thus 
leading to less reliable data in $\sigma_2$ and $\Gamma$, which in our case might be particularly the case for the measurements on the rather 
thick SrRuO$_3$ sample, with its low overall THz transmitted signal, above 40 cm$^{-1}$
\end{thebibliography}
\end{document}